\documentclass[10pt]{article}%

\usepackage[a4paper,top=2.5cm,left=2cm,bottom=2.5cm,right=2cm]{geometry}
\usepackage{amsmath}
\usepackage{amssymb}
\usepackage{amsthm} 
\usepackage{mathabx} 
\usepackage{graphicx}
\usepackage{hyperref} 
\usepackage[authoryear,sort&compress]{natbib} 
\usepackage{subfigure}  
\usepackage{color}
\usepackage{ulem} 

\newcommand{\apj}{The Astrophysical Journal}

\newtheorem{theorem}{Theorem} 
\newtheorem{corollary}{Corollary} 

\theoremstyle{remark}

\usepackage{lineno}

\usepackage{fancyhdr}
\begin{document}
\title{Distribution in the Geometrically Growing System and Its Evolution}

\author{Kim Chol-jun  \\%
\small \textit{Department of Astronomy, Faculty of Physics, \textbf{Kim Il Sung} University, DPR Korea }
\\
\small postal code:+850       
\\
\small email address: cj.kim@ryongnamsan.edu.kp
}
\maketitle
\begin{abstract}
Recently, we developed a theory of a geometrically growing system. Here we show that the theory can explain some phenomena of power-law distribution including classical demographic and economic and novel pandemic instances, without introduction of delicate economic models but only on the statistical way. A convexity in the low-size part of the distribution is one peculiarity of the theory, which is absent in the power-law distribution. We found that the distribution of the geometrically growing system could have a trend to flatten in the evolution of the system so that the relative ratio of size within the system increases. The system can act as a reverse machine to covert a diffusion in parametric space to a concentration in the size distribution.

\end{abstract}

\paragraph{keywords:} power-law; firm size distribution; the COVID-19 pandemic ; population in city; spectral hardening;


{Significance: 
Most economic systems that seem to show the power law distribution are analyzed by Gibrat's model, alias a geometrically growing system, which seems to give the log-normal distribution. We showed that the system can give an asymptotic power law if the correlation between parameters is considered. In this paper we show the system can lead to the spectral hardening provided the diffusion, or the increment of variances, along with the growth of the system.}

\section{Introduction}\label{sec:intro}
First, we explain the problem and some terminology. A system is composed of members and we call each member an item. The item has a measurable property, which is called a size. The population in city and firm size can be regarded as sizes while city and firm stand for item within country, which is in turn the system. The power law, alias Zipf's law or the Pareto distribution, states that the probability of an item is inversely proportional of a power of the size of the item: $p(Z)=\frac{M}{Z^\gamma}$, where $p(Z)$ stands for the frequency of item of size $x$, $\gamma$ for exponent of power and $M$ for the normalization constant. 

Historically, \cite{Pareto1896} showed that the distribution for income follows the power law. \cite{Estoup1916} and \cite{Zipf1932} observed the power law in word frequency in a novel and \cite{Auerbach1913} and \cite{Zipf1949} indicated the law for the population size of city. Much diverse things show the power-law distribution, for reviewing which we can refer to many works \citep[e.g. see][]{Mitzenmacher2004, Newman2005}. In fact, the author was interested in the cosmic ray spectrum, which seems a typical power-law. \cite{Salpeter1955} had found that the mass distribution of stars follows the power law. 

Several generative models for the power-law distribution are proposed, which we can categorize into some groups. The first models are based on a preferential attachment or ``rich-get-richer'' process \citep{Yule1924, Simon1955, Barabasi1999}. The second ones pursue the scale invariance, which is a peculiarity of the power law distribution \citep{Bak1987, Sneppen1995}. The third ones begin with demanded optimization \citep{Mandelbrot1953}. And others composite models derive the power-law distribution from specially assumed elementary distributions of the parameters \citep{Gibrat1931, Miller1957, Gabaix1999, Reed2004}. Those models show many possibilities generating the power-law distribution. However, all those are based on special assumptions. Though a postulation is the start of logic, but it should better have generality. And the logic should better cover wider range of size in data.

\section{The formalism for the distribution in the geometrically growing system}\label{sec:formal}
Recently, we developed a theory of a geometrically growing system (GGS) \citep{Choljun2022} on the basis of statistically maximally plausible assumptions, i.e. the normality of distribution of parameters. If the size of each item in an system grows geometrically or proportionately, we call the system geometrically growing. A GGS can be modeled by 
\begin{linenomath} \begin{align}\label{eq:ansatz}
Z=(1+\alpha)^t Z_0,
\end{align} \end{linenomath}
where $Z$ is the size of an item in system, $\alpha$ is the growth rate (hereafter simply, growth), $t$ stands for the the age of growth\footnote{In \cite{Choljun2022} the age was denoted by $n$.} and $Z_0$ is an initial size of the item. 

We can assumed the normal distribution for not only $\alpha$ but also $t$, which is statistically maximally plausible. 
Here we can introduce a correlation $R$ between $\alpha$ and $t$ without loss of generality because the correlation can be given even in completely arbitrary configuration of $\alpha$ and $t$.\footnote{\label{ftnt:covidcor} The instance of the COVID-19 pandemic in \cite{Choljun2022} showed a systematic correlation: the countries that had later outbreak of the pandemic show relatively lower growth, i.e. the positive correlation is obtained, which might be because they could have a warning or preparation. 

\cite{Gabaix2004} indicated that in some decades large cities grow faster, but in other decades small cities grow faster. This implies the sign of the correlation between the age (assuming that older cities are greater) and the growth flips over casually. 
}
If the correlation is positive ($R>0$), then the log-size ($Y=\log{Z}$)
 at the upper limit can be approximate by
\begin{linenomath} \begin{align}\label{eq:Y3}
Y_3=(A x_3+B)^2 +C,
\end{align} \end{linenomath}
while if the correlation is zero or negative ($R\leq0$), it is approximated by\footnote{$F$ and $G$ are interchanged in comparison with \cite{Choljun2022}.}
\begin{linenomath} \begin{align}\label{eq:Y4}
Y_4=F x_4+G,
\end{align} \end{linenomath}
where $x_3, x_4$ are variables following the standard normal distribution $N(0,1)$ and, if $\mu_t, \sigma_t, \mu_{\alpha}, \sigma_\alpha, \mu_i$ and $\sigma_i$ stand for the means and standard deviations of $t,\alpha$ and $Y_0=\log{Z_0}$, the parameters are given as follows:
\begin{linenomath} \begin{align}\label{eq:ABC}
A=a, \qquad
B=\dfrac{\sqrt{4a^2b^2+d^2}}{2a}, \qquad
C=c-\texttt{sgn}(R)\frac{d^2}{4a^2},  
\end{align} \end{linenomath}
\begin{linenomath} \begin{align}\label{eq:FG}
F=\texttt{sgn}(R)(a^2+b^2)+c, \qquad
G=\sqrt{2a^4+4a^2b^2+d^2},  
\end{align} \end{linenomath}
where
\begin{linenomath} \begin{align}\label{eq:abcd}
a=\sqrt{\sigma_{\alpha} \sigma_t \vert R \vert}, \qquad
b=\dfrac{\vert \mu_{\alpha} \sigma_t + \texttt{sgn}(R)\mu_t \sigma_{\alpha} \vert}{2\sqrt{\sigma_{\alpha} \sigma_t}}, \qquad
c=\mu_i+\mu_t \mu_{\alpha}-\texttt{sgn}(R)b^2, \notag \\
d=\sqrt{\sigma_i^2+(\mu_t^2\sigma_{\alpha}^2+\mu_{\alpha}^2\sigma_t^2)(1-\vert R \vert)+\sigma_t^2 \sigma_{\alpha}^2(1-R^2)},
\end{align} \end{linenomath}
and $\texttt{sgn}(R)$ stands for the sign of $R$ and $\vert\cdot\vert$ for the absolute value.

Thus, if $R>0$, the log-size behaves such as a $\chi^2$ variable while for the case of $R\leq0$, as a normal variable. We can derive the probability density function (PDF) of size for both case:
\begin{linenomath} \begin{align}
p_{Z_3}(Z)&=\frac{\exp\left[-\dfrac{1}{2} \left(\frac{\sqrt{(\log Z-C)}-B}{A} \right)^2 \right]}{2\sqrt{2\pi}A Z\sqrt{(\log Z-C)}}\qquad&\text{for }R>0,\label{eq:Z3PDF}\\
p_{Z_4}(Z)&=\frac{1}{\sqrt{2\pi}F Z}\exp\left[-\dfrac{\left(\log Z-G\right)^2}{2F^2}\right]\qquad&\text{for }R\leq0.\label{eq:Z4PDF}
\end{align} \end{linenomath}
We call $p_{Z_3}(Z)$ the log-completely squared chi ($\chi$) distribution with 1 degree of freedom (shortly, log-CS$\chi_1$ or log-CS) while $p_{Z_4}(Z)$ is well-known log-normal distribution. What is interesting is that the asymptotic exponent, i.e. the asymptotic slope in log-log scale diagram of the PDF, of the log-CS tends toward a constant:
\begin{linenomath} \begin{align}\label{eq:gamma3lim}
\gamma_{3\infty}=\lim_{\footnotesize \begin{subarray}{1} Y\to +\infty (Z\to +\infty) \text{ in } R>0  \end{subarray} } \frac{d(\log(p_{Z_3}(Z)))}{d(\log Z)}=-\left(1+\dfrac{1}{2A^2}\right)=-\left(1+\dfrac{1}{2\sigma_{\alpha}\sigma_t R}\right),
\end{align} \end{linenomath}
which says that the log-CS has asymptotic power-law $p(Z)=\frac{M}{Z^{\gamma_{3\infty}}}$ behavior. Especially, the asymptotic exponent depends only on the variances of the age and the growth. By the way, the exponent is negative so that we usually consider only its absolute value.

For the log-normal distribution, local slope is determined by the variance (more exactly, the standard deviation) $F$, which in turn depends on the means and variances of the parameters. 
 
\section{Statistics of the COVID-19 pandemic and its evolution}\label{sec:covid}

The propagation of the pandemic can be considered as a typical example of the geometrically growing system. In spite of seasonal rise and falls, the lock-down measures, administration of vaccines and appearance of variants, the propagation of the pandemic has been accelerated over 2 years since outbreak. The power-law distribution of infected in countries were reported in the early stage when the pandemic was propagating between countries \citep{Blasius2020,Beare2020}. However, once propagation between countries had been saturated, the distribution should be deviated from the power-law.

\begin{figure*}
\centering
\subfigure[]{\includegraphics[width=0.42\textwidth]{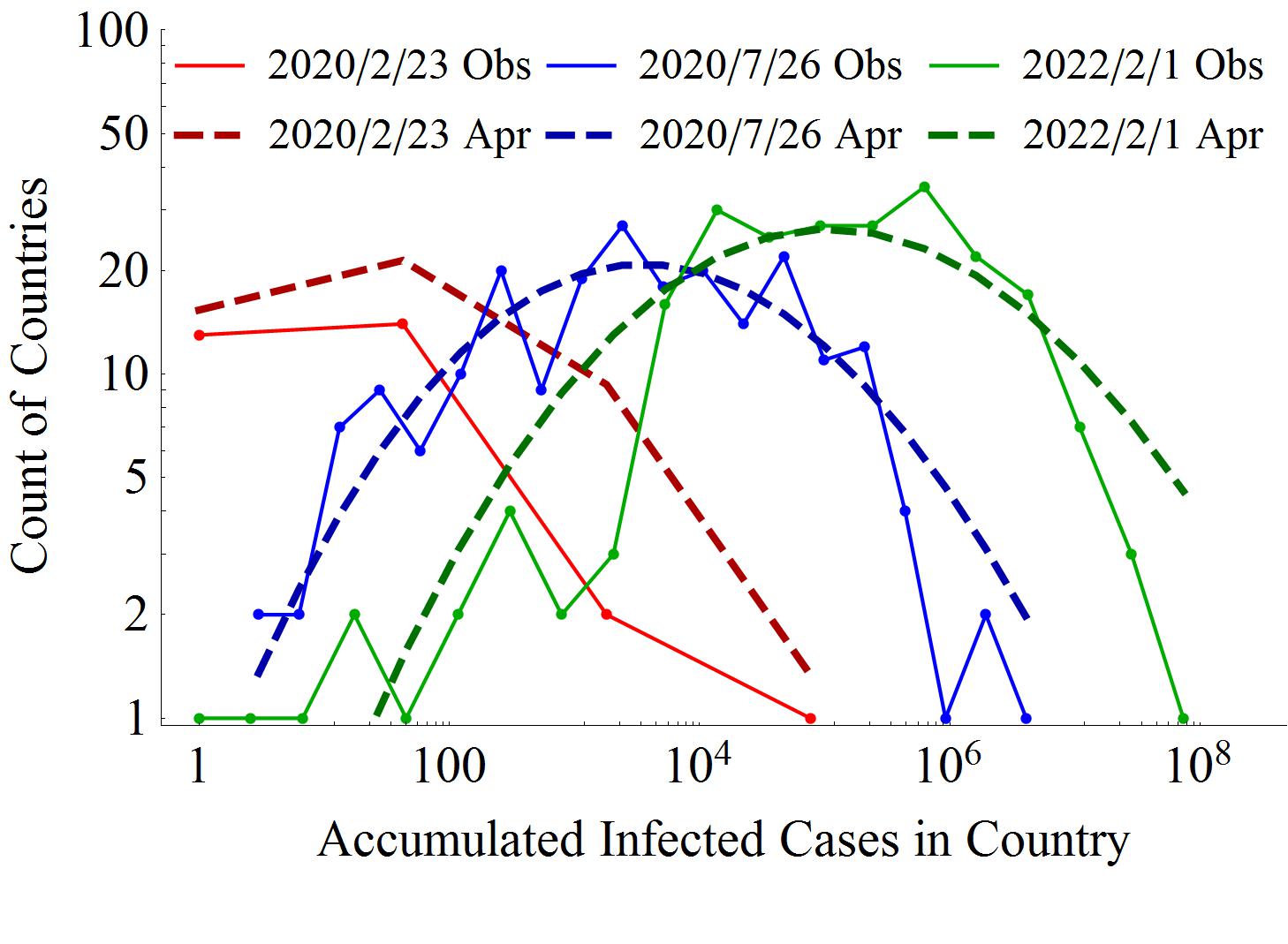}\label{fig:COVIDEvolCount}}  
\subfigure[]{\includegraphics[width=0.42\textwidth]{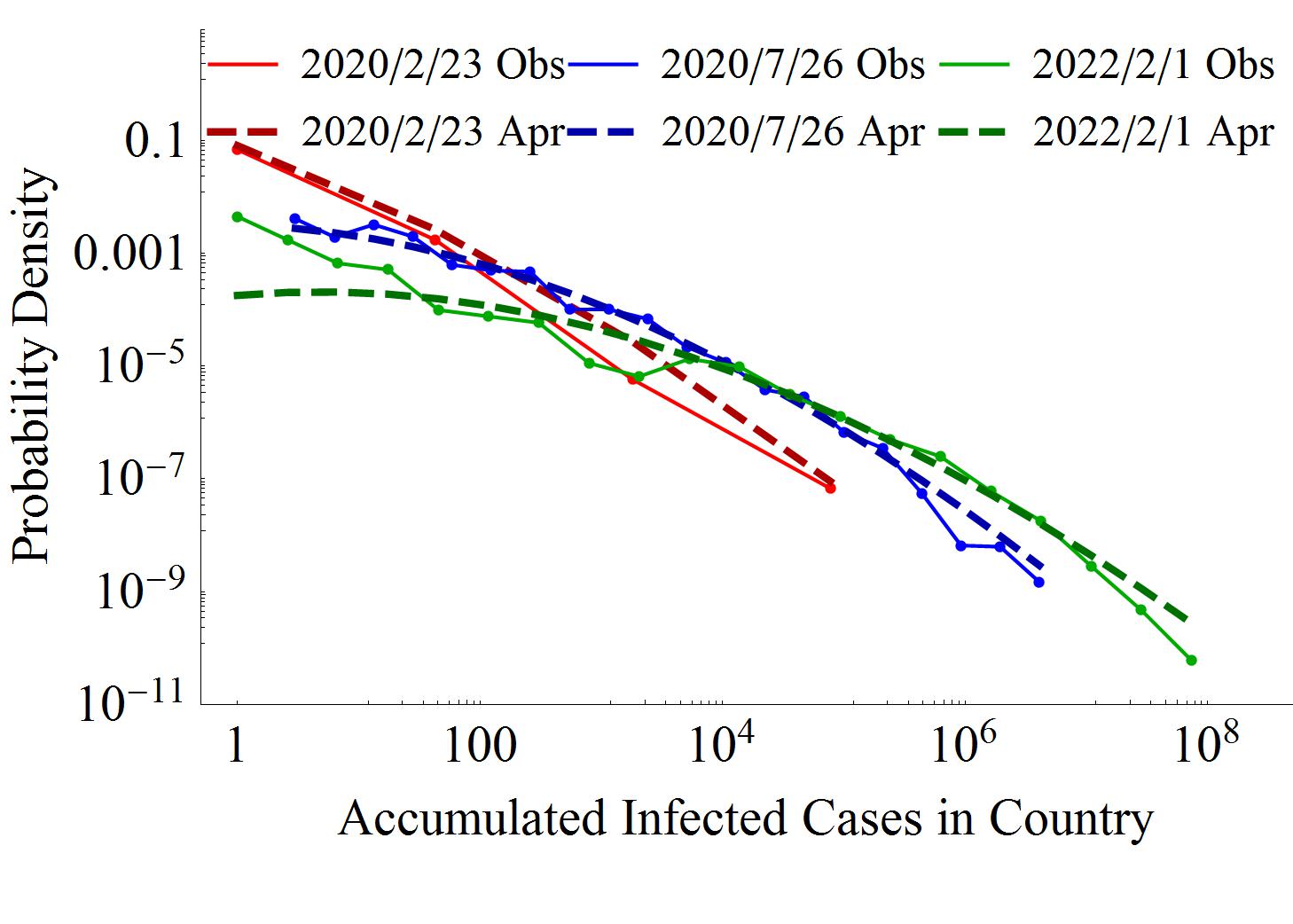}\label{fig:COVIDEvolPDF}}
\subfigure[]{\includegraphics[width=0.3\textwidth]{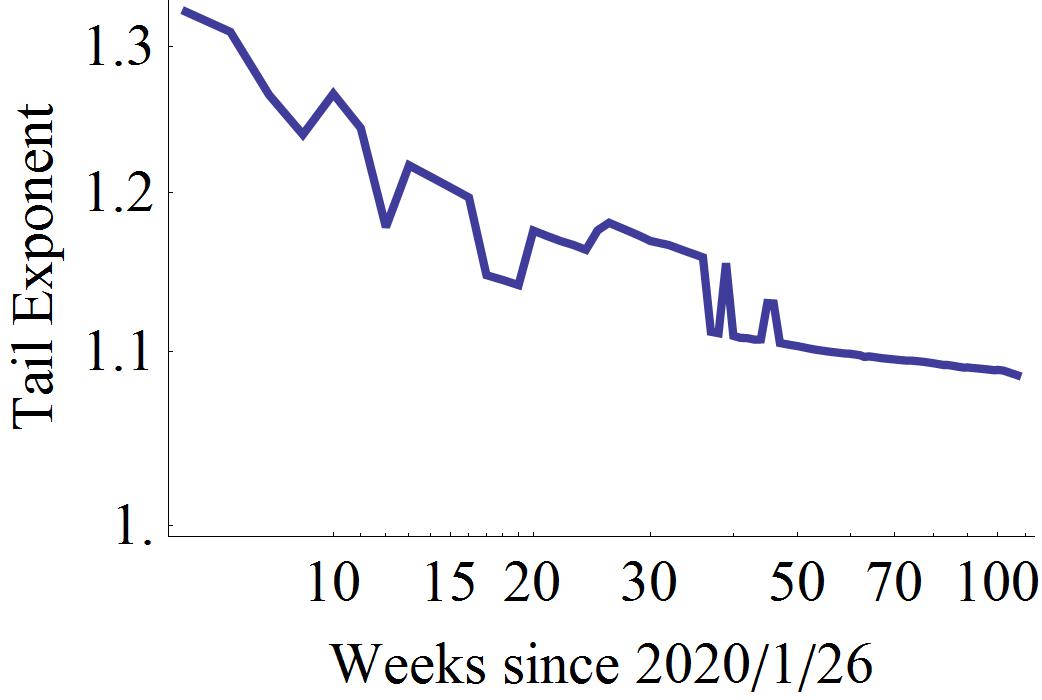}\label{fig:COVIDEvolExp}}\qquad \qquad \qquad
\subfigure[]{\includegraphics[width=0.3\textwidth]{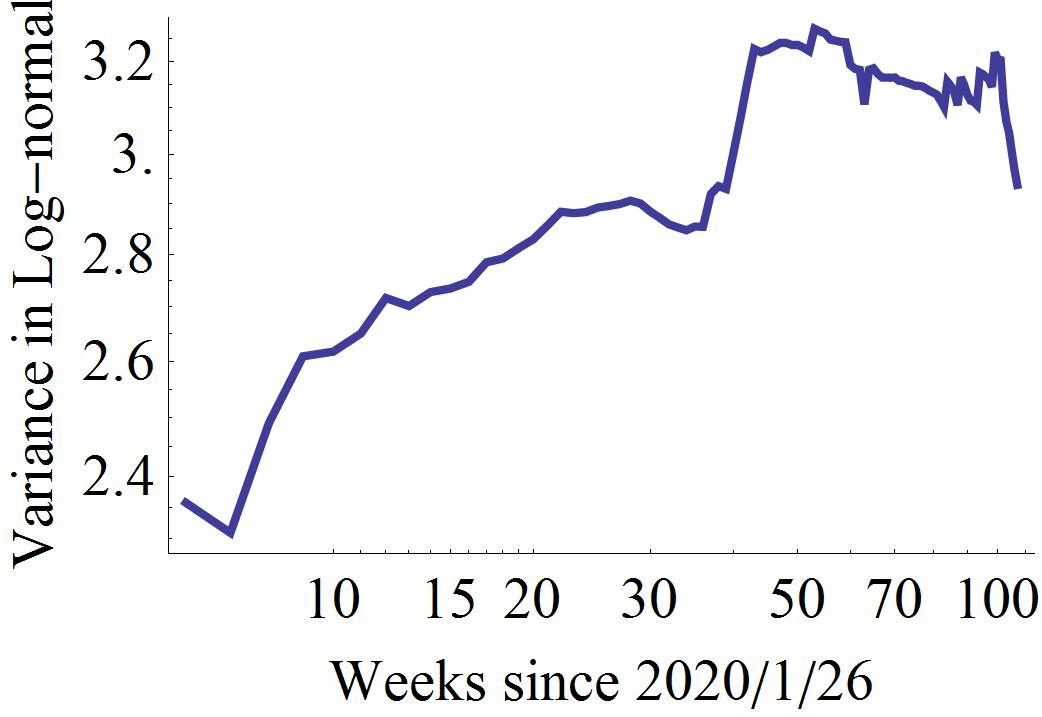}\label{fig:COVIDEvolVar}} 
\caption{\label{fig:COVIDEvol} The distribution of the accumulated infected cases of the COVID-19 pandemic in countries and its evolution. The distribution at several stages of propagation and the log-CS$\chi_1$ approximation in (a) count histogram and (b) the probability density. The evolution of (c) the tail exponent of the distribution in probability (see (b)) and (d) the variance in log-normal approximation. 
} 
\end{figure*}

\cite{Choljun2022} showed that the distribution of accumulated infected in countries in May 2021 could be approximated by the log-CS excellently. Note that the approximation is not a best-fitting but derived from the history of the pandemic. In fact, distributions of the age and growth are similar to the normal and their correlation turned out to be positive.$^{\ref{ftnt:covidcor}}$ Figure~\ref{fig:COVIDEvol} shows the consistency between observation and the log-CS approximation at several stages of the pandemic: as examples, in late February 2020 (the early stage), late July 2020 (the saturation of propagation between countries) and early February 2022 (the propagation of the Omicron variant).\footnote{ Data for COVID-19 propagation is available, for example, at the website of Our World in Data \url{https://ourworldindata.org/coronavirus/}.
} Especially for July 2020, the observation curve is weaving around the log-CS approximation. Distribution in early February 2022 seems to be getting distorted by the unprecedented quick propagation of the Omicron variant. 



In the above count and probability histograms we could hardly sense a change of the slope of the probability curve. 
If we try to use a maximum likelihood (ML) estimation of the power-law exponent \citep{Newman2005}
\begin{linenomath} \begin{align}\label{eq:MLtailExp}
\hat{\gamma}=1+N\left(\sum_{i=1}^{N} \log\frac{z_i}{z_{\text{min}}}\right)^{-1},  
\end{align} \end{linenomath}
where $z_i$ stands for the size for data items and $N$ for the number of the items, we should set $z_{\text{min}}$, i.e. the lowest allowable size or truncation size. However, the distribution is not the power-low over all domain of size but only in the tail part of big size. We could set $z_{\text{min}}$ as the modal (most probable) size in the probability histogram (Fig.~\ref{fig:COVIDEvolPDF}) and determine the tail exponent. Figure~\ref{fig:COVIDEvolExp} shows that the tail exponent appears to decrease over all the past time in spite of local rises. 

We could propose another proxy for the tail exponent: the variance of the log-normal distribution. In fact, the distribution can be approximated by the log-normal (Eq.~\ref{eq:Z4PDF}) as well. This distribution has not an asymptotic exponent and local slope is determined by the variance $F$ (Eq.~\ref{eq:FG}). We can estimate this variance also in the maximum likelihood approach:
\begin{linenomath} \begin{align}\label{eq:MLFG}
\hat{F}^2=\frac{1}{N}\sum_{i=1}^{N} (\log z_i-\hat{G})^2, \qquad
\hat{G}=\frac{1}{N}\sum_{i=1}^{N} \log z_i.
\end{align} \end{linenomath}
This approach has advantage over the above evaluation of the tail exponent because the selection of an optimal truncation size or $z_{\text{min}}$ does not matter. 
In both the log-CS and log-normal distributions the slope increases for bigger size so that the tail exponent could be evaluated greater for greater size and vice versa even though the distribution still remains the same. The greater variance corresponds to the smaller local slope or the tail exponent. In the evolution of the COVID-19 pandemic, the variance seems to increase (Fig.~\ref{fig:COVIDEvolVar}), which is coincident with decreasing the tail exponent as aforementioned.

\section{Statistics for population in city and in country}\label{sec:citypop}

The distribution of population in city is a classical instance of the power-law. The growth of population over centuries shows an exponential or geometrical profile though sometimes was so saturated that expressed by the logistic function. 
Therefore, we can express the evolution of population by the geometrically growing system (GGS). 

\begin{figure*}
\centering
\subfigure[]{\includegraphics[width=0.3\textwidth]{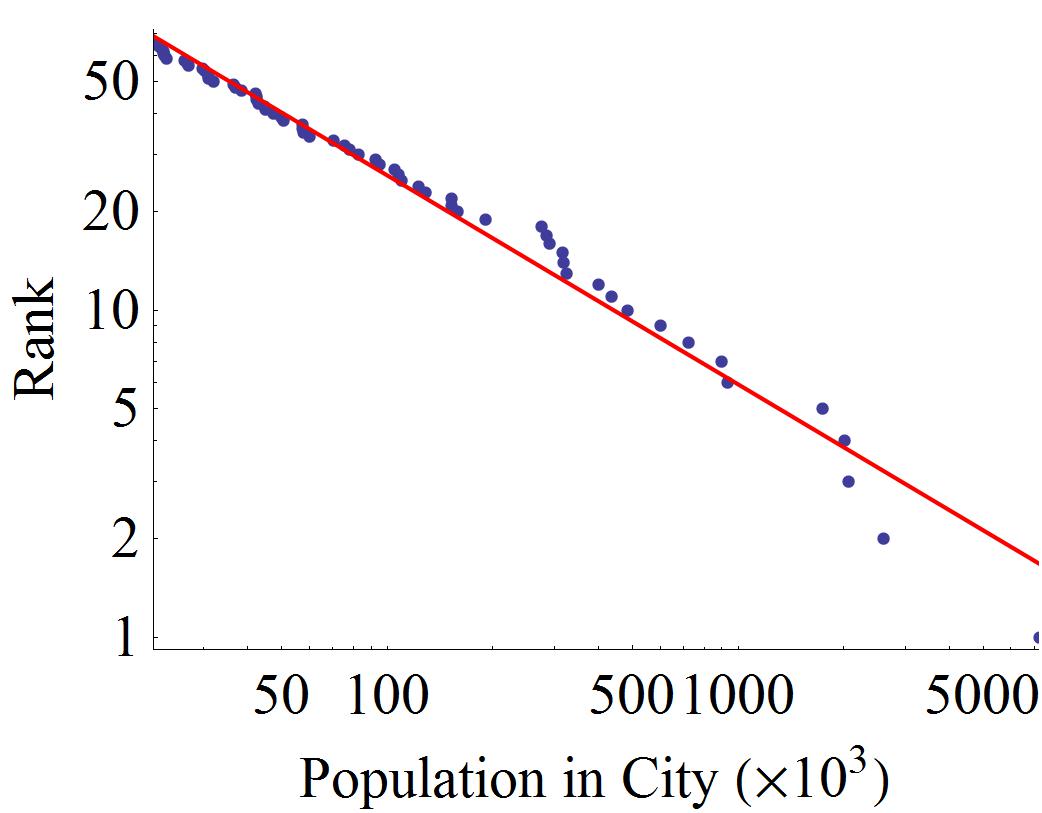}\label{fig:CityPopNgramUS}}  
\subfigure[]{\includegraphics[width=0.35\textwidth]{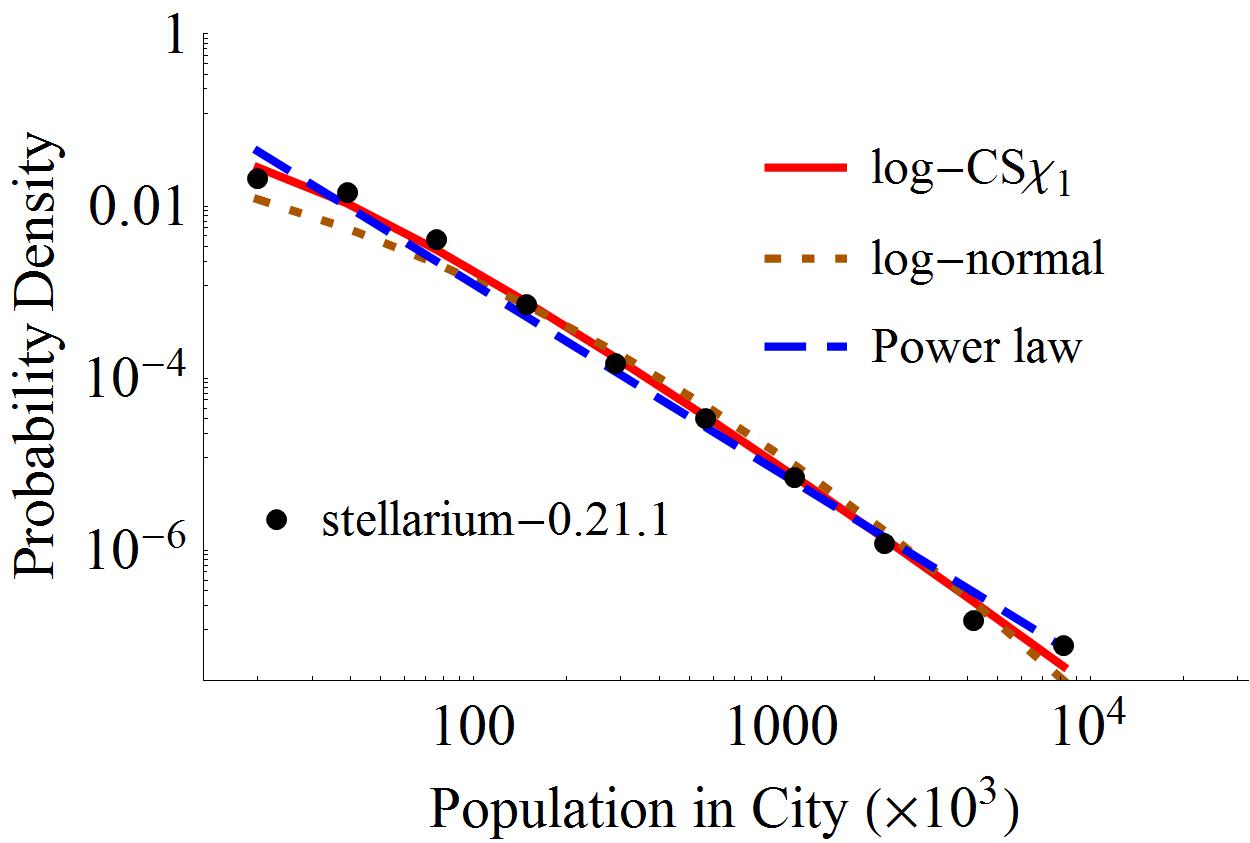}\label{fig:CityPopUS}}
\subfigure[]{\includegraphics[width=0.35\textwidth]{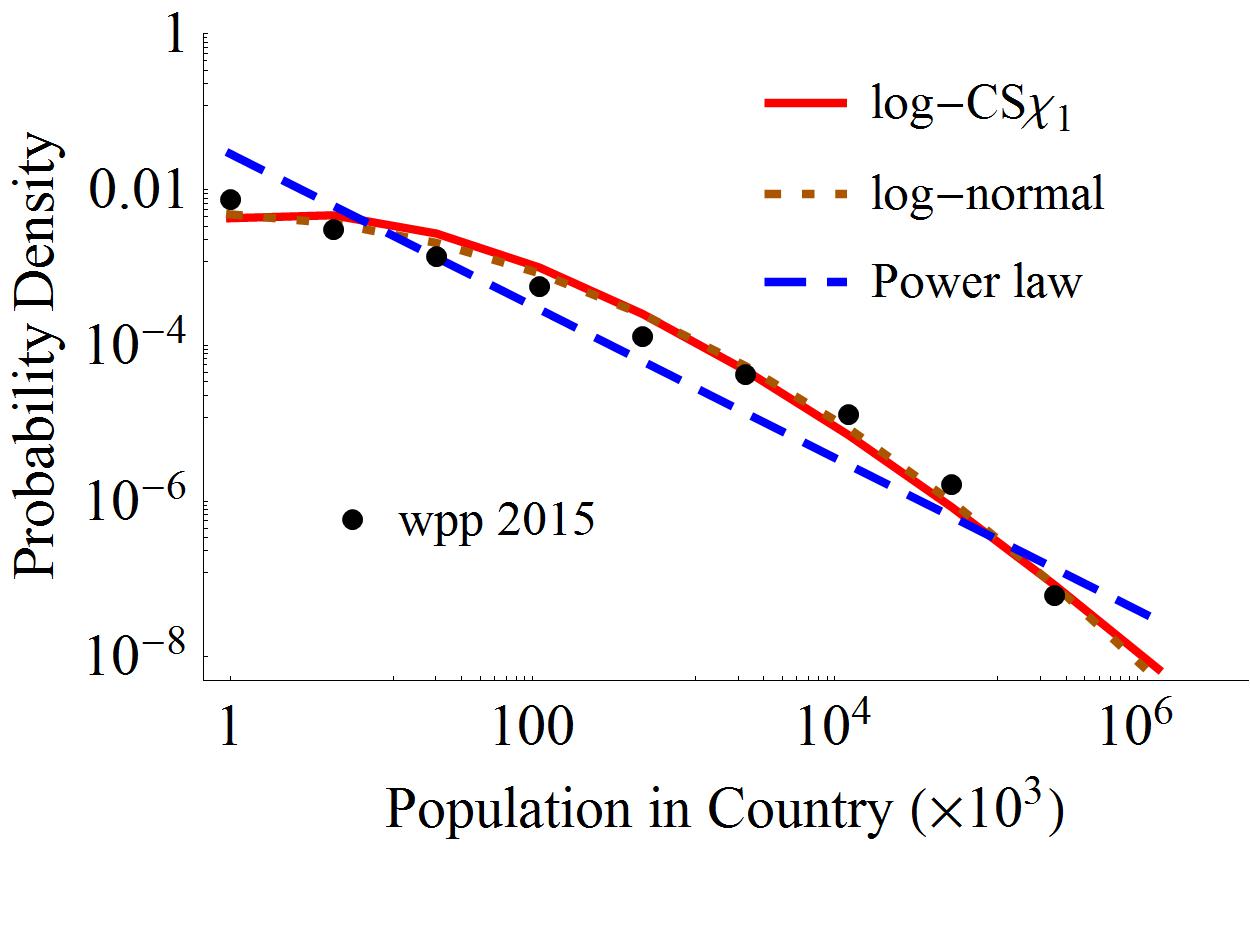}\label{fig:wpp2015}}
\subfigure[]{\includegraphics[width=0.3\textwidth]{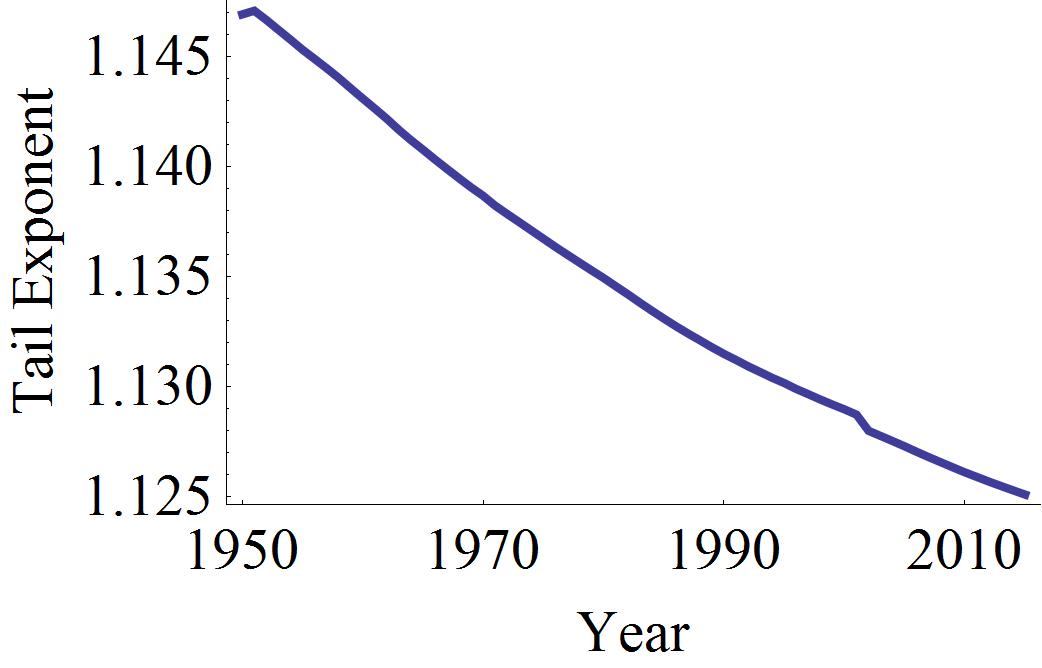}\label{fig:wppEvolExp}}
\subfigure[]{\includegraphics[width=0.3\textwidth]{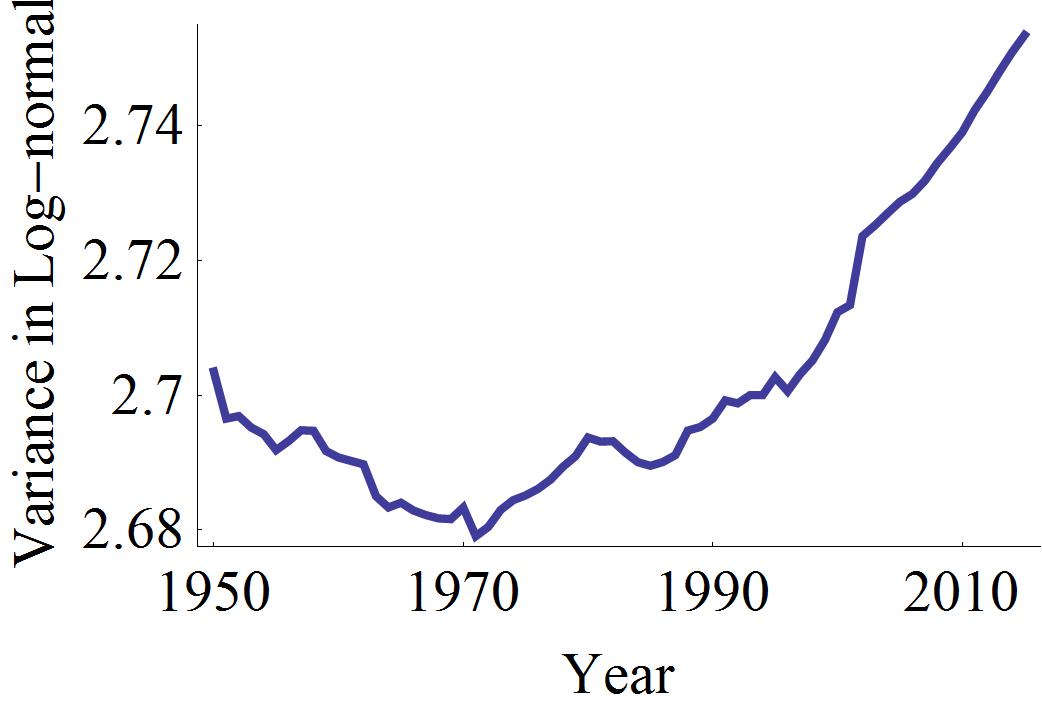}\label{fig:wppEvolVar}} 
\caption{\label{fig:popEvol} The distribution of population in city and country and its evolution. (a) A plot of log(Rank) vs. log(Population) for cities of U.S. The linear regression gives log(Rank) = 6.1946 - 0.6387 log(Population). (b) The bet-fitting to the distribution of population in cities of U.S. The log-CS appears the closest. (c) The population in countries and areas over the world in 2015. The evolution of (d) the tail exponent and (e) the variance in the log-normal approximation for population in countries and areas over the world.} 
\end{figure*}

First, we analyzed the population in city within a country using data in stellarium-0.21.1.\footnote{We use the dataset for cities compiled as observation locations on the globe in stellarium-0.21.1, which is an open-source astronomical software. The software is available at Stellarium Github webpage \url{https://github.com/stellarium/stellarium/releases/}. 
The dataset covers $\sim$24,000 cities with their location, population and other information gathered between 2006 and 2019. Because of discontinuity in lower population, we limit cities to the population over 20,000.}
Analyzing the population in cities for U.S., \cite{Gabaix1999} indicated that the populations in biggest cities follow the power-law distribution and Zipf exponent is almost unity. We obtained Zipf exponent 1.29 for U.S. (Fig.~\ref{fig:CityPopNgramUS}), which differs from 1 according to \cite{Gabaix1999} and 0.639 for Iraq, for example. We perform the best-fitting analysis for population in cities of U.S. with various approximations: the log-CS, the log-normal and power-law (Fig.~\ref{fig:CityPopUS}). We infer the best-fit parameters by a Markov Chain Monte Carlo (MCMC) method, especially making use of the Metropolis-Hastings algorithm \citep{Metropolis1953, Hastings1970}. 
$R^2$ with best-fit parameters is evaluated: $R^2=0.9956$ for the  log-CS, $R^2=0.9905$ for the power-law and $R^2=0.9833$ for the log-normal approximation. Therefore, we can prefer the log-CS as the closest approximation in considering population in city within country. 
\cite{Ioannides2013} claimed that most cities in the U.S. obeys a log-normal, but the upper tail and therefore most of the population obeys a power-law. In fact, the log-CS and log-normal distributions are almost indistinguishable except for in the infinity \citep{Choljun2022}. 

Next, we perform the best-fitting analysis for the distribution of population in countries and areas over the world, making use of World Population Prospects (WPP) 2015 dataset\footnote{The data is available at the website of World Population Prospects \url{https://population.un.org/wpp/}.} with the approximations (Fig.~\ref{fig:wpp2015}). With the best-fit parameters inferred by the MCMC method, $R^2$ is evaluated: $R^2=0.9851$ for the  log-CS, $R^2=0.9380$ for the power-law and $R^2=0.9920$ for the log-normal approximation. Therefore, in this case we can prefer the log-normal as the closest approximation.

\begin{figure*}
\centering
\subfigure[]{\includegraphics[width=0.3\textwidth]{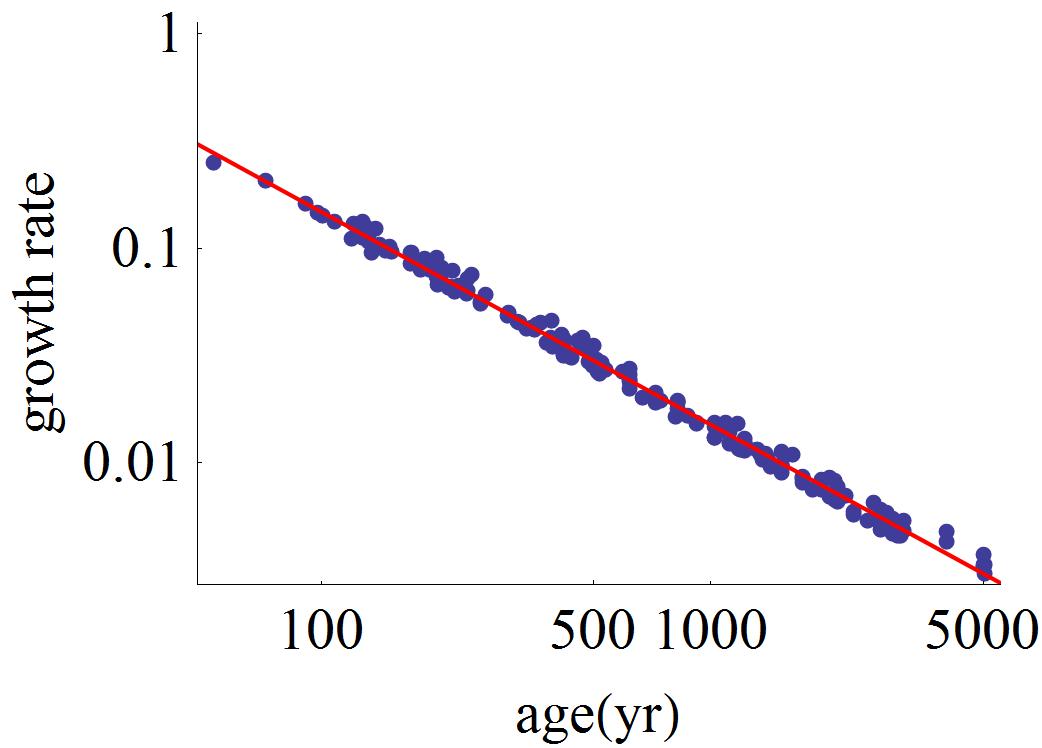}\label{fig:wppCor}} 
\subfigure[]{\includegraphics[width=0.3\textwidth]{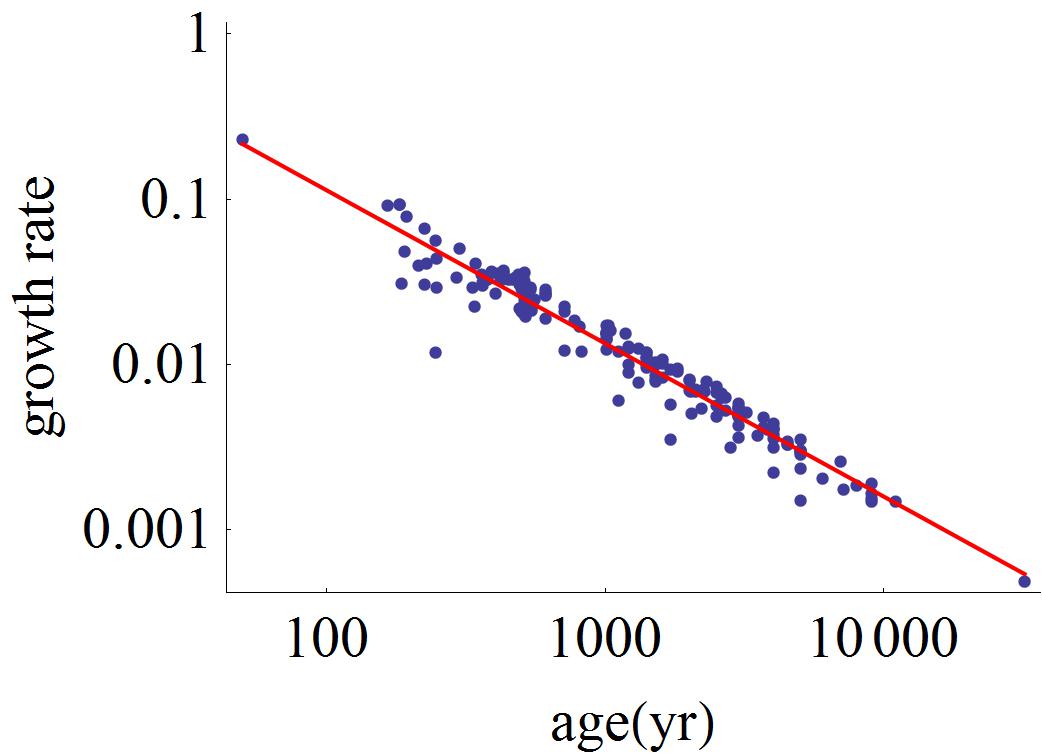}\label{fig:BriCor}}\\
\subfigure[]{\includegraphics[width=0.3\textwidth]{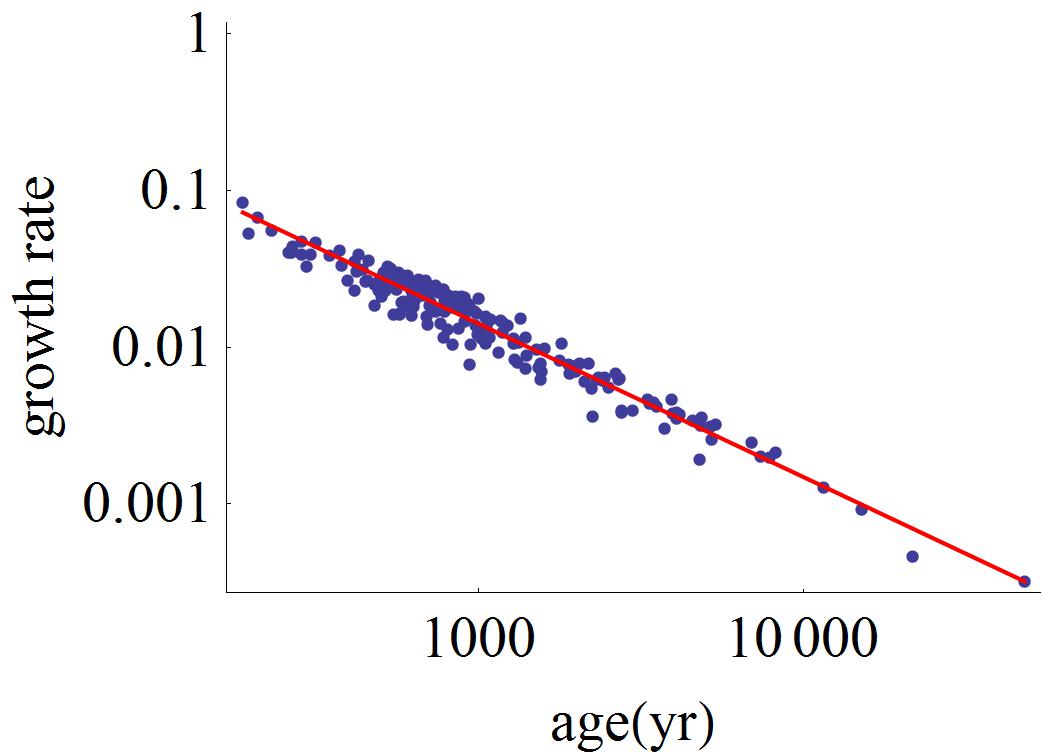}\label{fig:ExtCor}}
\subfigure[]{\includegraphics[width=0.3\textwidth]{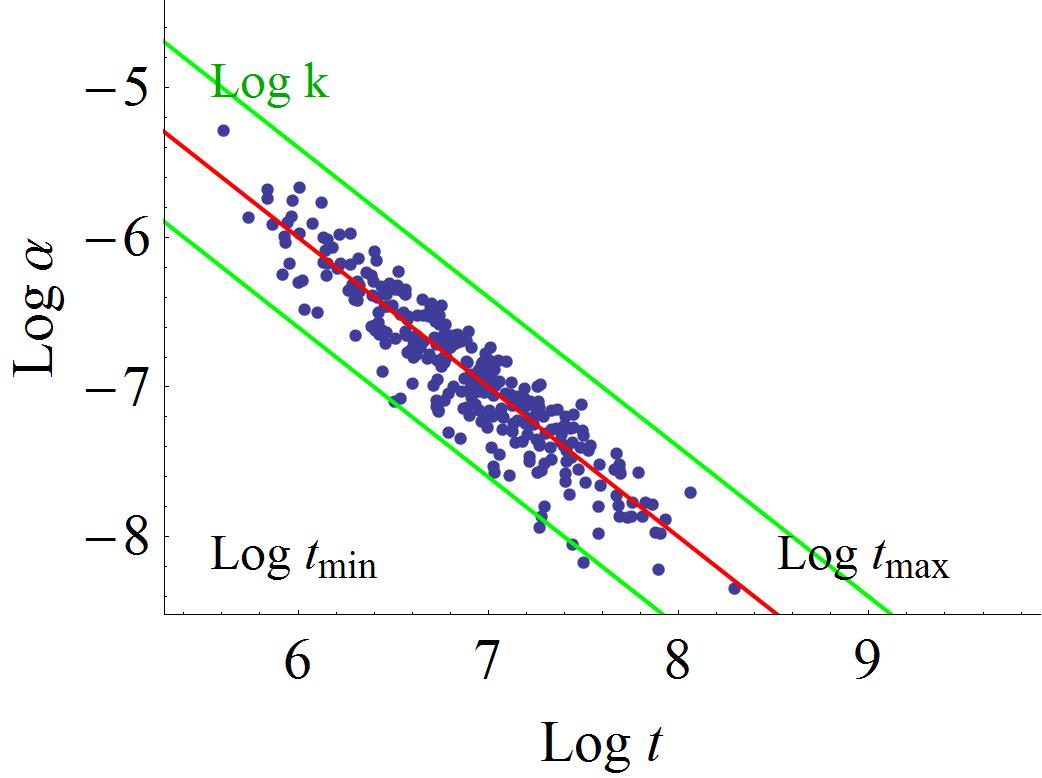}\label{fig:TheoCor}}
\caption{\label{fig:worldPopCor} The relation of the age and the growth for population in countries and areas over the world assuming the extreme initial condition. (a) The relation between a ``administration age'' and the corresponding growth. The slope is $-0.9943$ in log-log scale and the correlation between the age and the growth is $R=-0.6191$. (b) For a ``habitation age,'' the slope is $-0.9281$ and the correlation is $R=-0.3599$. (c) For an ``extrapolation age,'' the slope is $-0.9771$ and the correlation is $R=-0.3911$. (d) The difference of size forms a stripe which corresponds to a negative correlation. } 
\end{figure*}

On the other hand, we try to apply our approach of the GSS. We expect the correlation between the age and the growth$^{\ref{ftnt:covidcor}}$. For that, we have inspected the age of countries in the Korean Great Encyclopedia\footnote{the Korean Great Encyclopedia, Pyongyang, Encyclopedia Press, 2001}. We date the starting epoch of country by appearance of the first administration, e.g. the first dynasty or city-state, the independence and so on. However, for many African and American countries, we should consider that the establishment of the colony had changed greatly the composition of population in those countries. We are afraid that we might take missed or distorted official record of real history for many countries and areas in extracting typical dates. The growth rate is so evaluated for each country that the population was originated from a couple of Adam and Eve, which approximation was applied to COVID-19 pandemic in \cite{Choljun2022}. Surprisingly, the growth rate and age show a exactly inverse relation, furthermore its exponent is almost unity (Fig.~\ref{fig:wppCor}). This gives a negative correlation between the age and growth and, of course, we could expect that the world population should follow the log-normal distribution.

Also from Britanica\footnote{Encyclop$\ae$dia Britannica Ultimate. Reference Suite. Chicago: Encyclop$\ae$dia Britannica, 2014.} we extracted the date of the first habitation of the tribe or immigration. This kind of age that could be called ``habitation age'' are much older than the previous ``administration age.'' But the negative correlation is obtained still (Fig.~\ref{fig:BriCor}). We also considered another kind of age which is extrapolated from the current growth of population: the origin of age is set so that the initial population was also a couple. We call such an age ``extrapolation age.'' Figure~\ref{fig:ExtCor} shows relation between the age and the growth those are obtained by the extrapolation from the period 1950-2015 of WPP dataset. Also a negative correlation is given. And, for any kind of age, the inverse relation between the age and the growth still holds and their exponent are near unity.

We analyze this fact. From Eq.~\eqref{eq:ansatz} we can derive 
\begin{linenomath} \begin{align}\label{eq:lnansatz}
\log{z}-\log{z_0}=t\log(1+\alpha)\approx t\cdot\alpha.
\end{align} \end{linenomath}
If $t\alpha=b$ and $b$ is determined to extent of $k$ times, then $t$ or $\alpha$ are also determined to extent of $k$ times and in log-log diagram of $t$ vs. $\alpha$ appears a stripe of width $\log k$ (Fig.~\ref{fig:TheoCor}). This stripe has slope of $-1$ surely so that a negative correlation between $\alpha$ and $t$ is obtained. In order that a positive correlation gets, this stripe must cover all the range of $t$ in dataset. Therefore, it must hold that $k\geqslant\sqrt{\frac{t_{\text{max}}}{t_{\text{min}}}}$, which we can rewrite from Eq.~\eqref{eq:lnansatz}:
\begin{linenomath} \begin{align}\label{eq:reqposcorr}
\frac{\left(\log z-\log z_0\right)_{\text{max}}}{\left(\log z-\log z_0\right)_{\text{min}}}\geqslant\frac{t_{\text{max}}}{t_{\text{min}}},
\end{align} \end{linenomath}
where $t_{\text{max}}$ and $t_{\text{min}}$ could be appropriate extremes of the dataset, e.g. of $3\sigma$ region. This might be a necessary condition for positive correlation between $\alpha$ and $t$. To summarize, we can claim a theorem. 
\begin{theorem} \label{th:theo1}
The geometrically growing system can have positive correlation between the growth and the age only if Eq.~\eqref{eq:reqposcorr} satisfies.
\end{theorem}
From the theorem we could derive another conclusion:
\begin{corollary} \label{th:cor1}
If new items with the lowest size are continuously born within the geometrically growing system, the system should be approximated by the log-normal. On the other hand, the system could be approximated by the log-CS after the creation of new items with lowest size has been stopped.
\end{corollary}
For the countries over the world, if we take an initial condition $z_0=2$ (a couple) and consider that the maximum and minimum population in countries are now $z_{\text{max}}=10^9, z_{\text{min}}=10^3$ and $t_{\text{max}}=5000$ and $t_{\text{min}}=50$, then we obtain $100$ in r.h.s and only $3$ in l.h.s. of Eq.~\eqref{eq:reqposcorr}. However, if we would consider a non-constant initial condition (in real circumstance), we might gain so much greater value of l.h.s that we could perform log-CS approximation. For the case of the COVID-19 pandemic, $t_{\text{max}}-t_{\text{min}}\approx120$ and $z_{\text{min}}=1,z_0=1$, so we can apply the log-CS soon after saturation of propagation between countries. Note that Eq.~\eqref{eq:reqposcorr} is not a sufficient condition.

We inspect the evolution of the population distribution over the world. WPP 2015 dataset provides the population in countries and areas in time 1950-2015. As aforementioned, the upper slope of the distribution can be evaluated by both methods: the tail exponent and the variance in the log-normal approximation. Figures~\ref{fig:wppEvolExp} and \ref{fig:wppEvolVar} show the similar trend of the flattening slope as in case of the COVID-19 pandemic. This means that the variance of the parameters such as the age and the growth is so growing that the variance of the size distribution grows, and the tail exponent decreases. 

\begin{figure*}
\centering
\subfigure[]{\includegraphics[width=0.35\textwidth]{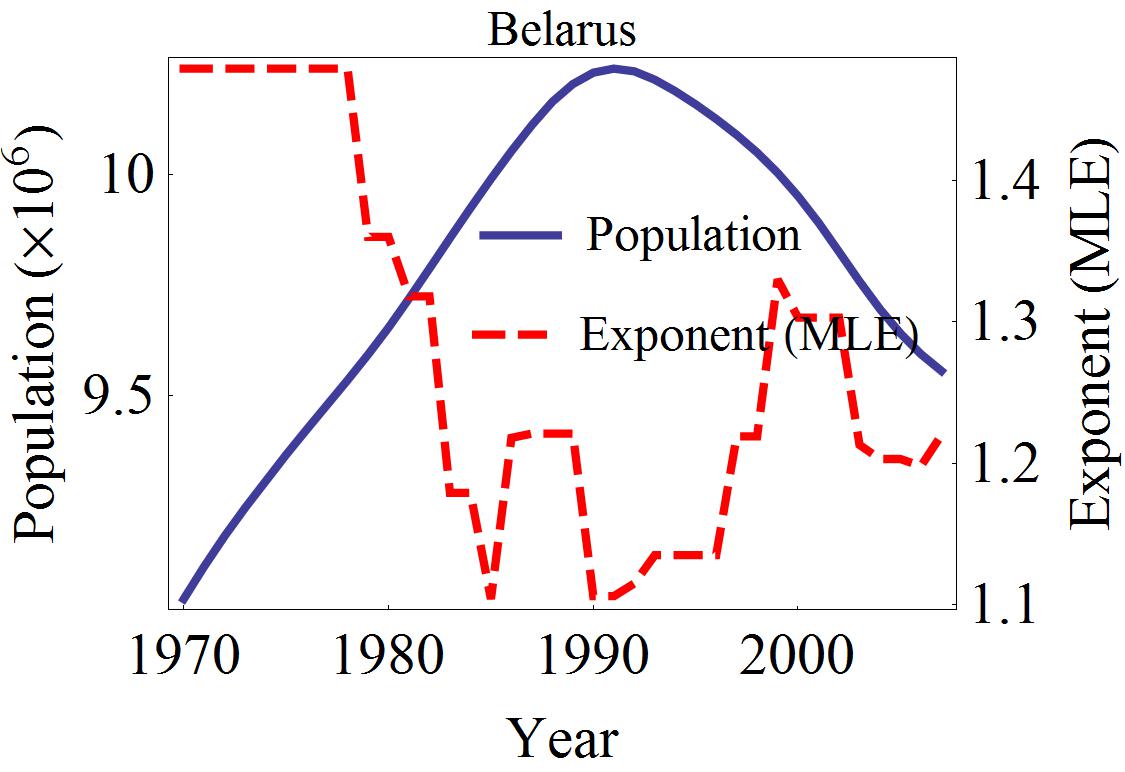}\label{fig:BelarusEvol}}  
\subfigure[]{\includegraphics[width=0.35\textwidth]{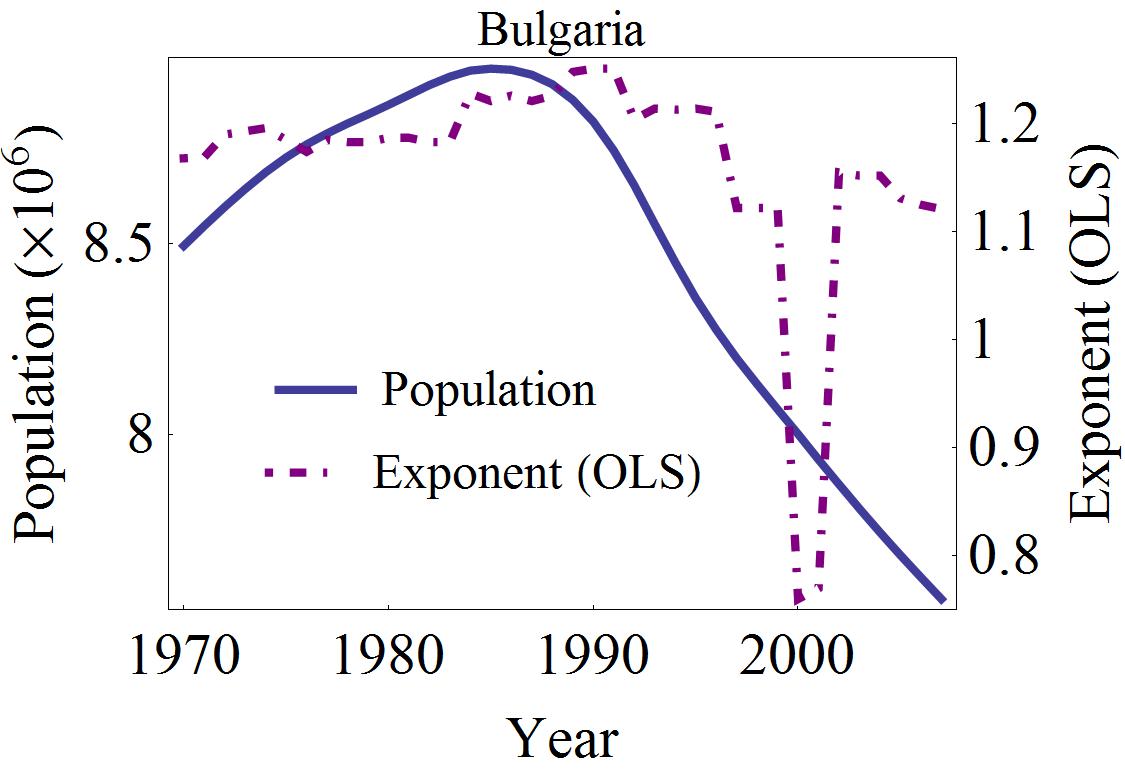}\label{fig:BulgariaEvol}}
\subfigure[]{\includegraphics[width=0.35\textwidth]{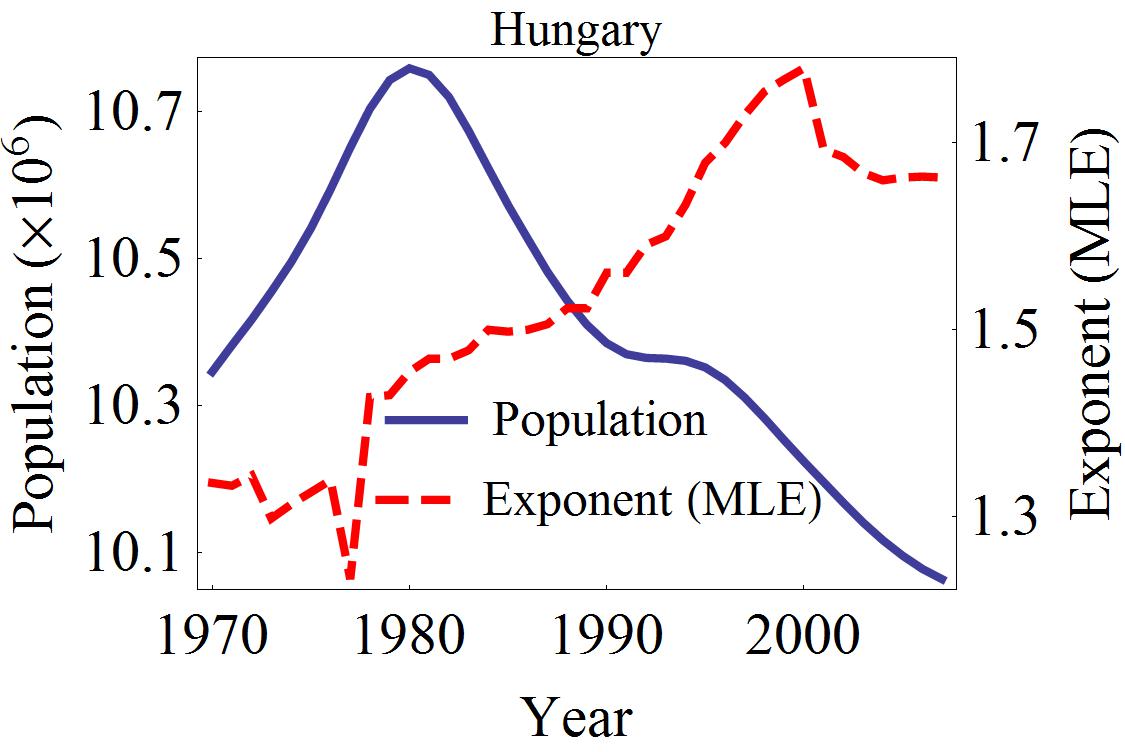}\label{fig:HungaryEvol}} 
\subfigure[]{\includegraphics[width=0.35\textwidth]{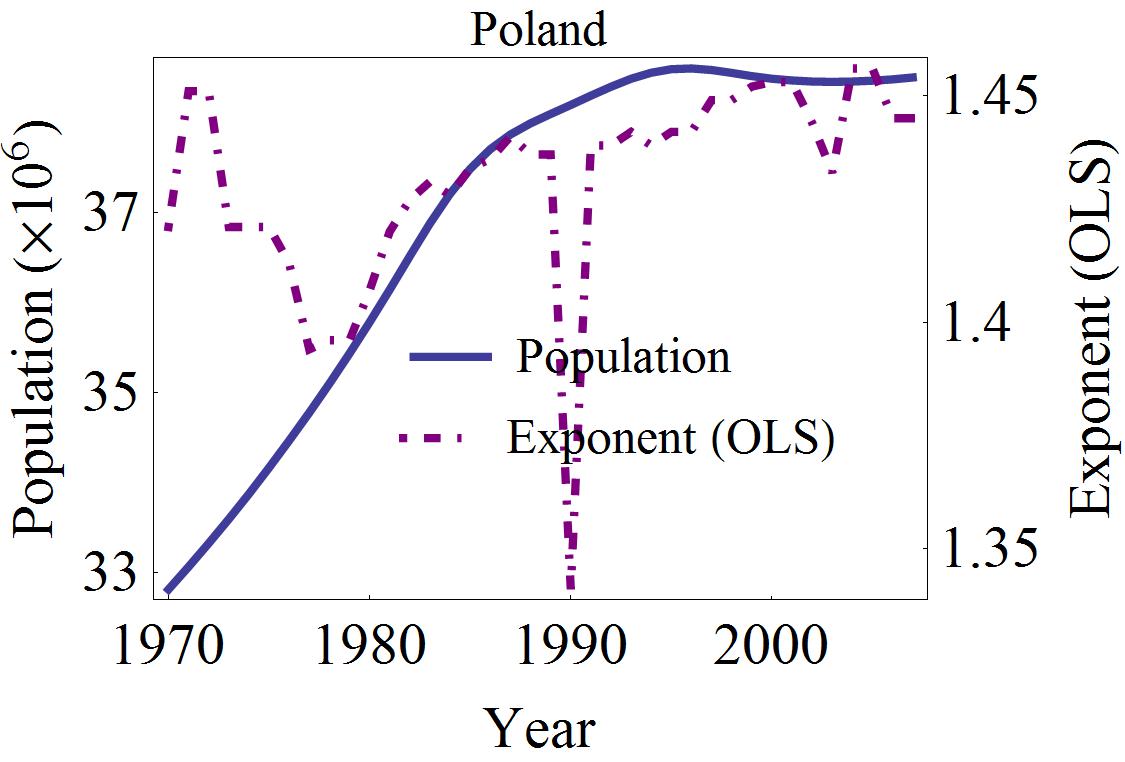}\label{fig:PolandEvol}}
\subfigure[]{\includegraphics[width=0.35\textwidth]{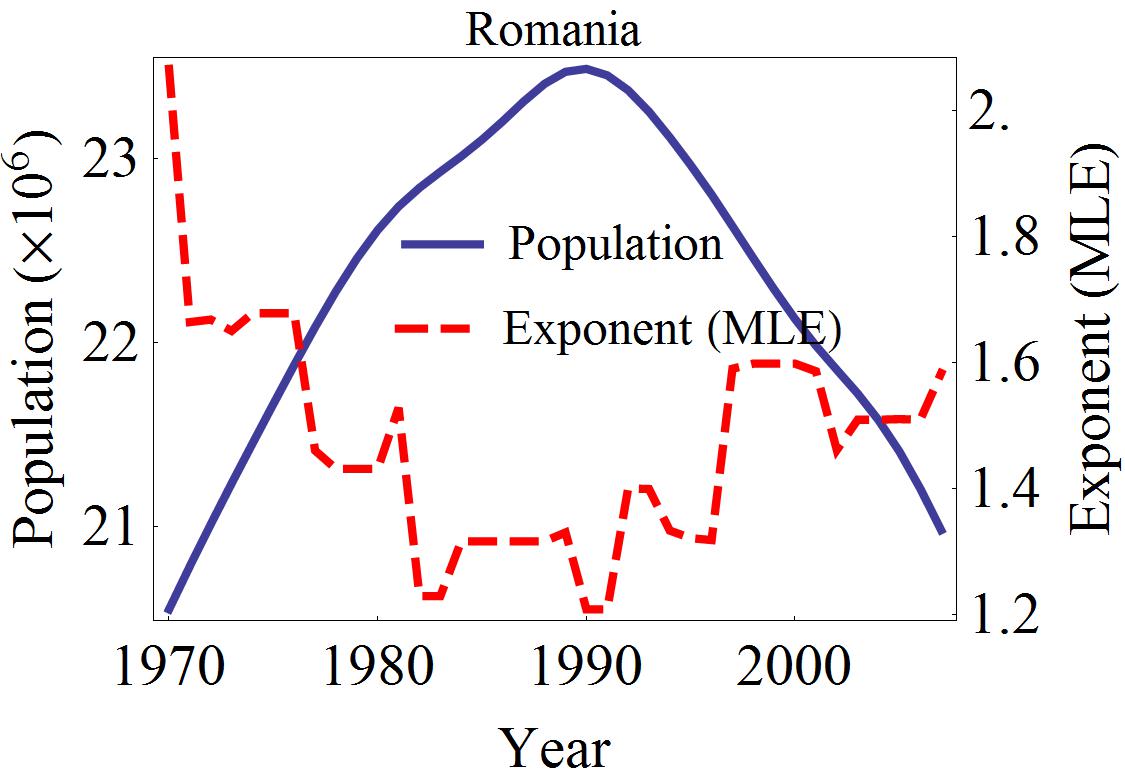}\label{fig:RomaniaEvol}}
\subfigure[]{\includegraphics[width=0.35\textwidth]{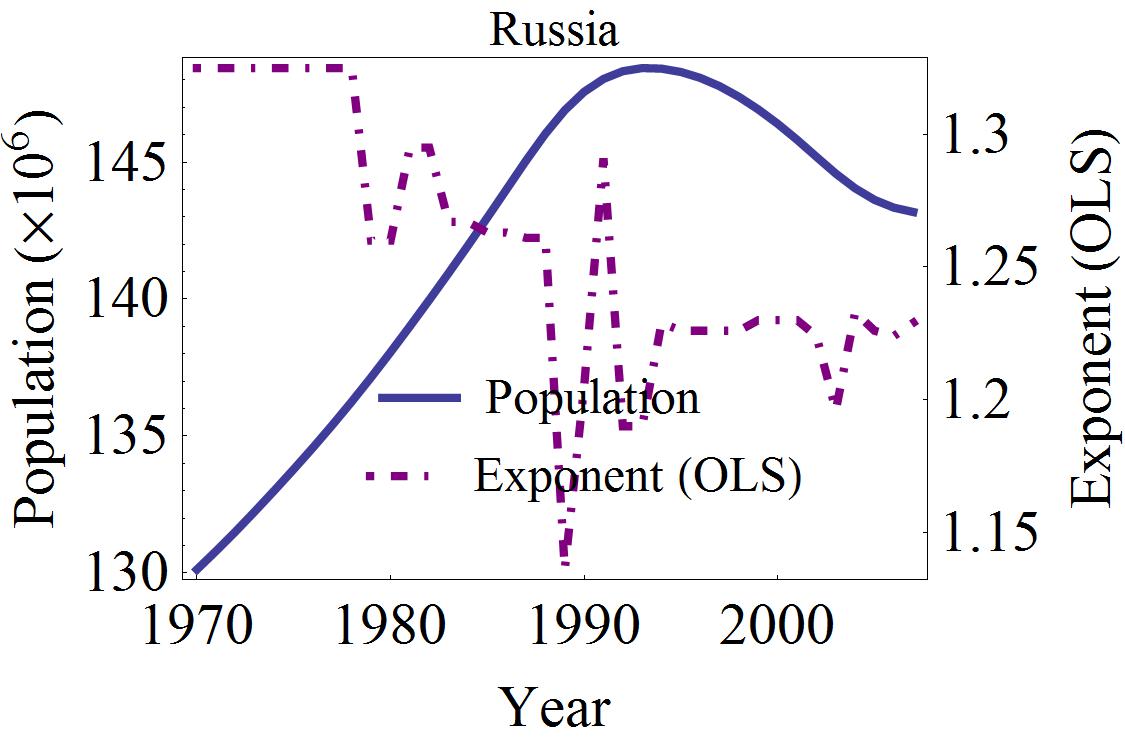}\label{fig:RussiaEvol}}
\subfigure[]{\includegraphics[width=0.35\textwidth]{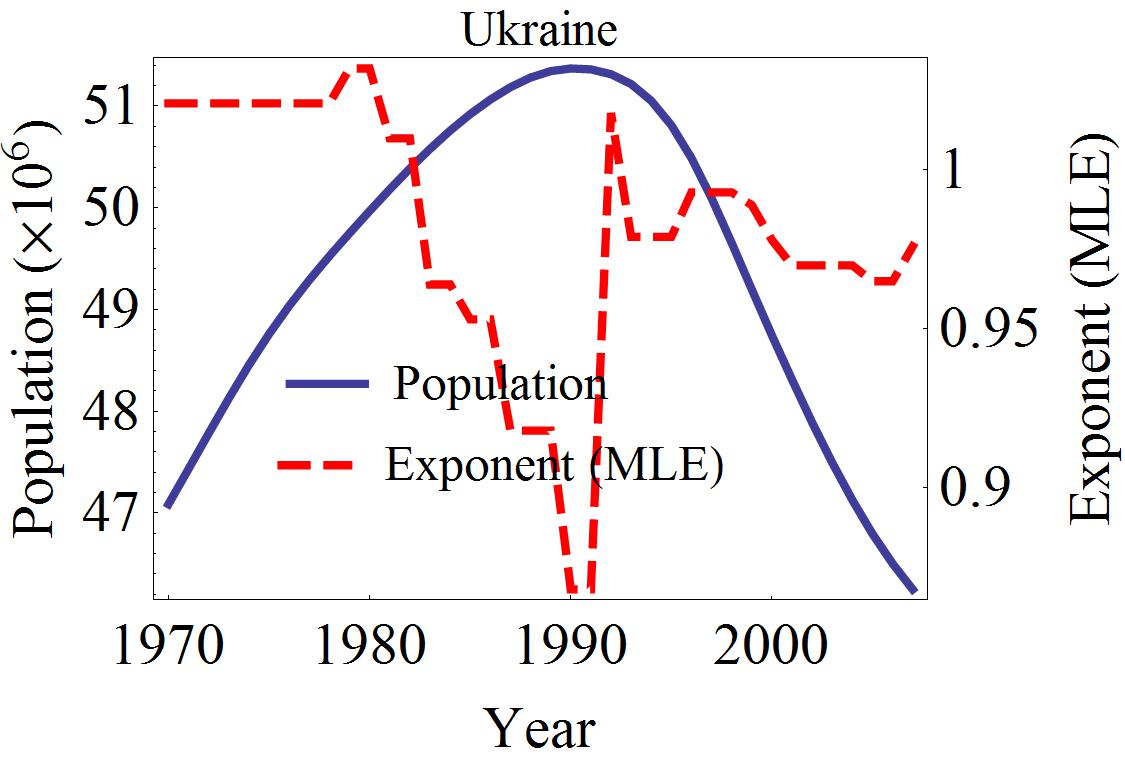}\label{fig:UkraineEvol}}
\subfigure[]{\includegraphics[width=0.35\textwidth]{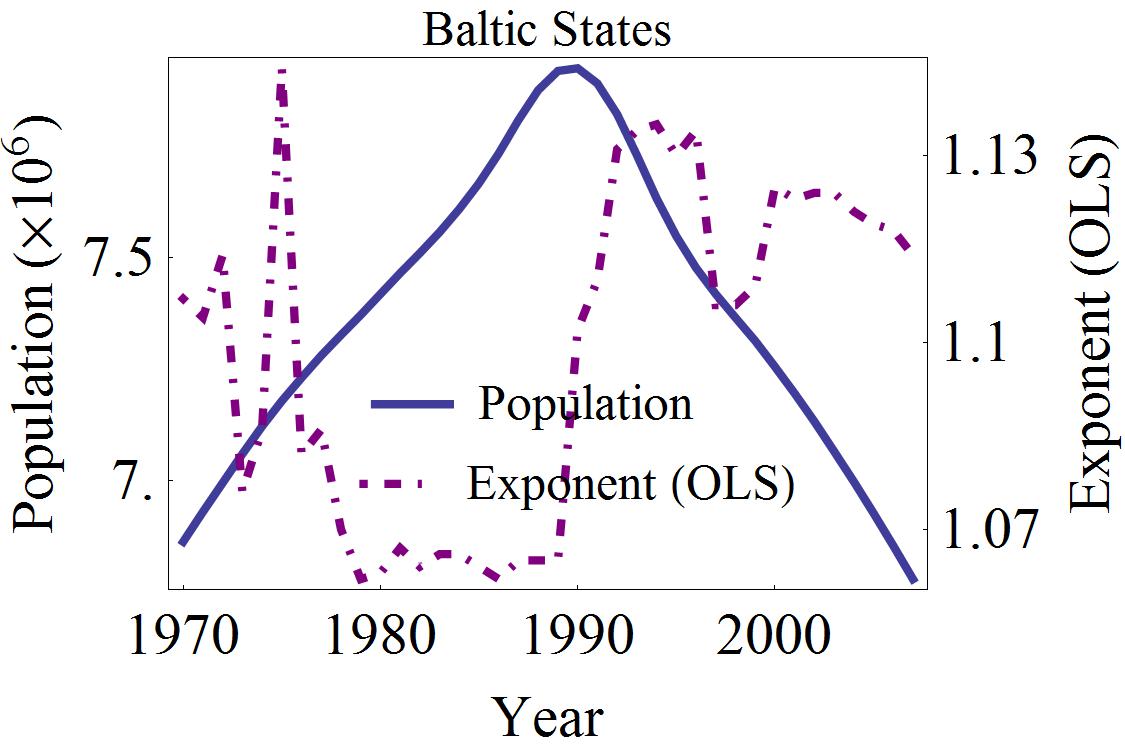}\label{fig:BalticEvol}}
\caption{\label{fig:CEEEvol} Evolution of the population and the tail exponent for CEE countries. (a) Belarus, (b) Bulgaria, (c) Hungary, (d) Poland, (e) Romania, (f) Russia, (g) Ukraine, (h) Baltic states (Estonia, Latvia and Lithuania). The data is extracted in \cite{Necula2010} where the exponent is calculated by two methods: ordinary least squares (OLS) estimation for linear regression and maximum likelihood estimation (MLE). For some countries one of either estimations of the exponent shows abrupt rise and fall so that we neglect. For most countries except Bulgaria and Poland we can observe inverse relation between the population and the tail exponent in evolution.} 
\end{figure*}

The tail exponent for population in city within a country also evolves. Citing previous works, \cite{Gabaix2004} indicated that the tail exponent for the U.S. decreased in the period from 1900 to 1990 to imply a greater concentration. \cite{Gonzales2010} assured a monotonically decreasing of the tail exponent with time, provided the truncation number of cities keeps as 10,000. Interestingly, observing data for dynamics of cities in the central and eastern Europe (CEE) countries during 1970-2007 \citep{Necula2010}, we can find that in most countries the exponent has almost a negative relation with the population itself: if the population increases, the exponent decreases and vice versa (Fig.~\ref{fig:CEEEvol}). The exponent for European cities in the middle ages seems to decrease after 1500 \citep{Bairoch1988, Gonzales2019}, and only then Zipf’s law was reported to emerge for cities in Europe \citep{Dittmar2011}. 

Considering countries, quite different with respect to wealth, size and geography, \cite{Pinto2012} claimed that the countries presenting higher wealth levels
reveal higher values of the exponent while most African countries unveil smaller values of the exponent. In our simulation, oldest Asian countries such as Iraq and China appear to have smallest exponent. However, their reasoning was so not obvious: if it was right, the exponent over the world should increase as the world economy proceeds but the exponent surely is decreasing. Our approach can give an obvious reasoning: younger countries might have greater exponent and vice versa. In fact, the most more developed countries are younger while the older countries are underdeveloped so that more wealthy countries could appear to have higher exponent, which, however, is casual but not inevitable. 

\cite{Gabaix2004} related the urbanization with the economic factor, e.g. the economic integration and the international trade. Then why is the exponent increasing in some countries in spite of economic progress? \cite{Necula2010} proposed political factor to determine the urbanization. We could give a statistical analysis ahead of or including all the economic or political or any other factors. For example, the exponent depends on the variances of the parameters.

\section{Statistics of firm size}\label{sec:firmsize}

The power-law distribution has appeared widely in economic and financial phenomena \citep[e.g. see][]{Farmer2008}. 
The power-law distribution in firm size that could be measured by diverse properties 
have anounced long ago \citep{Zipf1949, Ijiri1977}. Making use of Economic Census 1997, \cite{Axtell2001} showed the power-law distribution in firm size measured by employees and revenue. The firm size is a quantity which could be apt to grow geometrically. In fact, we commonly evaluate the growth of firm in terms of proportionality but not additivity. 

\begin{figure*}
\centering
\subfigure[]{\includegraphics[width=0.37\textwidth]{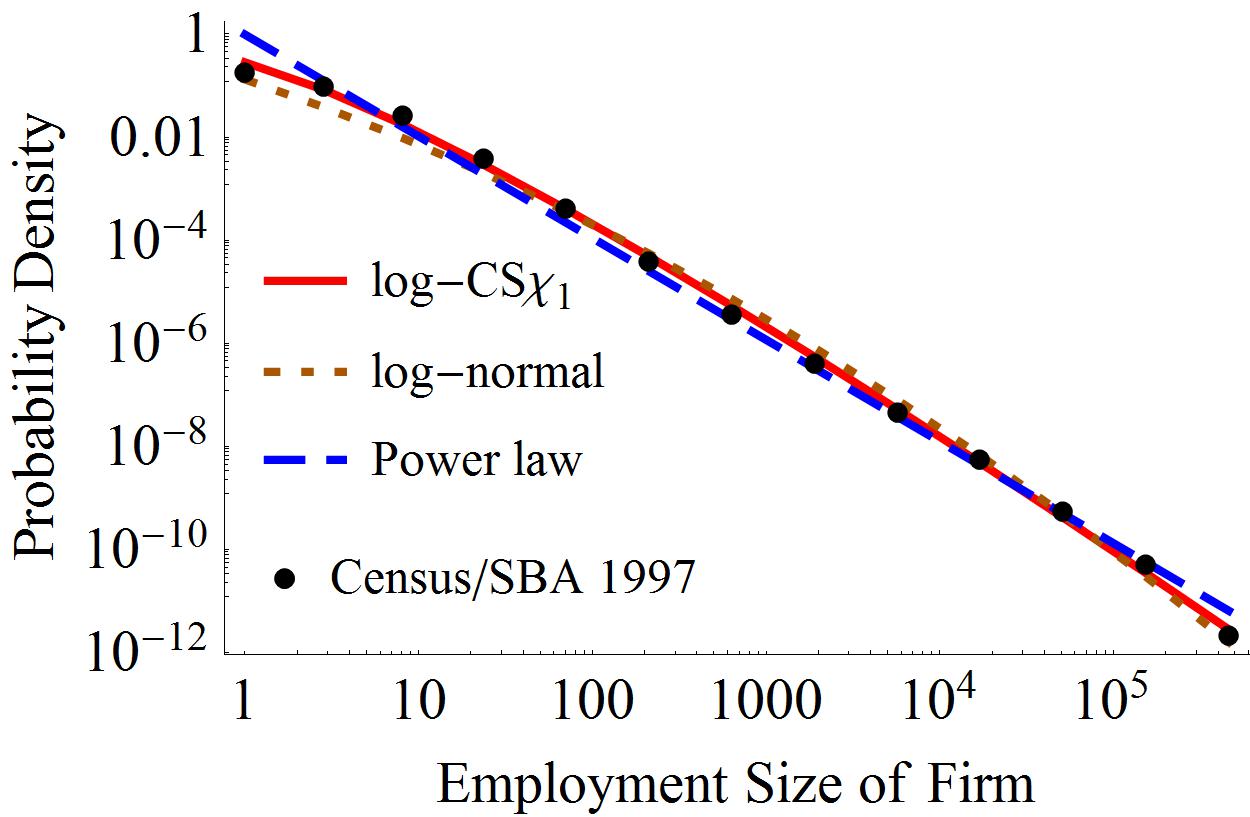}\label{fig:FirmSizeAxtell}}
\subfigure[]{\includegraphics[width=0.3\textwidth]{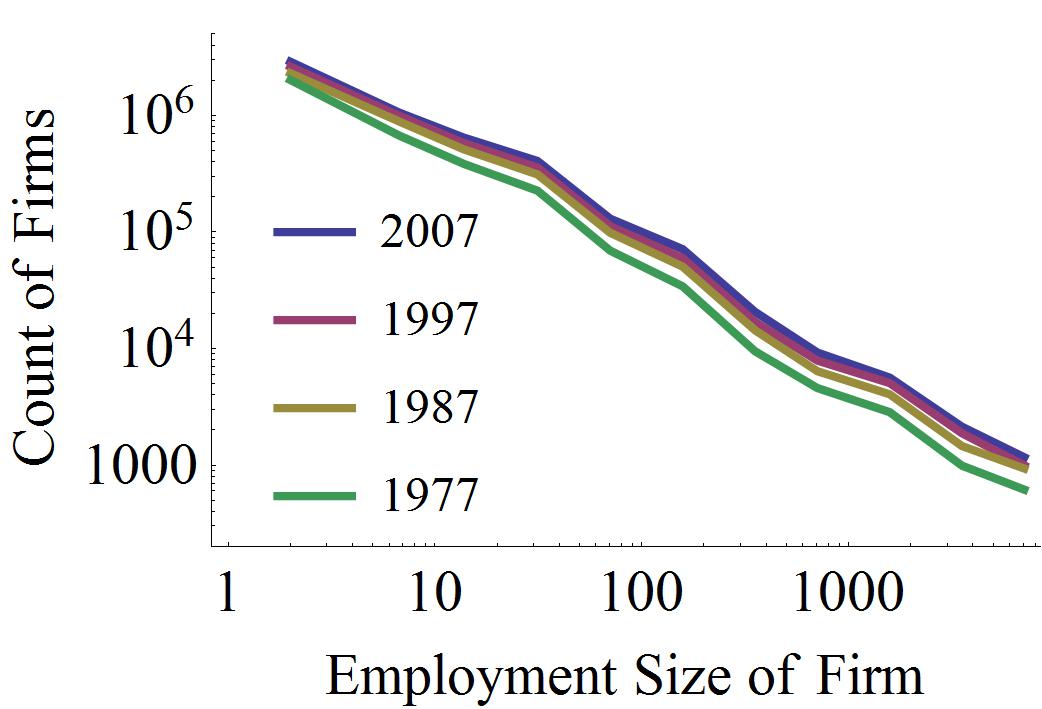}\label{fig:FirmSizeDistrEvolv}}
\subfigure[]{\includegraphics[width=0.3\textwidth]{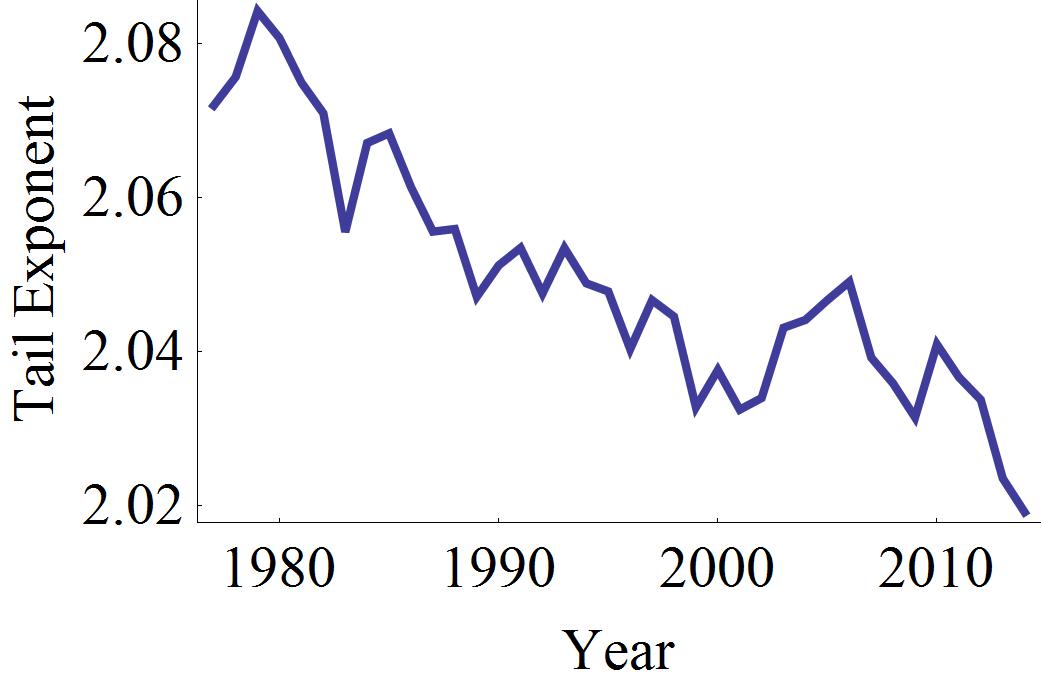}\label{fig:TailExponentEvolv}}   
\caption{\label{fig:FirmSize} Distribution of the employment size of firm and evolution of the tail exponent. (a) The log-CS$\chi_1$, log-normal and power-law fitting for distribution of the employment size of firm in Census/SBA 1997 dataset. The evolution of (b) histogram and (c) the tail exponent for firm's employment size in BDS 1977-2014 dataset. The tail exponent is evaluated not in count histogram such as (b) but in probability histogram such as (a) by linear regression.}
\end{figure*}

We analyse the data in \cite{Axtell2001}, where numerical data for the size of firm expressed by the number of employees (the employment size) were shown explicitly. 
In fact, the distribution has a convex form in low-size part, which is in more favor of the log-CS rather than the pure power-law modeling. 
\cite{Giovanni2010} analyzed French firms, and obtained a similar convex profile of distribution.
We neglect 0-size firms as Axtell did. 
We perform the best-fitting with the various approximations by the MCMC method (Fig.~\ref{fig:FirmSizeAxtell}). For the power-law fitting $R^2$ is obtained the same as Axtell: $R^2=0.9932$. The  log-CS fitting gives a greater value: $R^2=0.9987$. We should expect that the dataset could be approximated by the log-normal: $R^2=0.9952$. This says that the log-CS could be the closest to the real dataset.

As aforementioned, the geometrically growing system can be modeled alternatively by either log-CS and log-normal depending on the correlation between the growth and age. 
Data for employment dynamics by firm age, 1987-2005, from the Census Bureau Business Dynamics Statistic and Longitudinal Business Database, showed that young firms have higher employment growth rates, if they survive, than older firms \citep{Haltiwanger2009, Haltiwanger2010}. This might be because the growth of older or greater firms seems to be saturated due to market limitation while this limitation does not affect younger firms so the latter appears to have higher growth in spite of higher establishment exit. 
Analyzing data from the EFIGE survey that sampled French, Italian and Spanish firms in the period from 2001 to 2008, \cite{Navaretti2012} showed that younger firms have a highly probability of experiencing high growth rates both in the short-run (e.g. for 1-year) and in the long-run (i.e. for existing age). Therefore, we could expect a negative correlation between firm age and firm growth. This should lead to the log-normal fitting to the distribution, though the real dataset seems closer to the log-CS. This might be originated from non-normal distribution of the age and growth.

We trace the evolution of the distribution. We inspect data from the Census Bureau Business Dynamics Statistic (BDS), 1977-2014.\footnote{The data is available at website of Small Business Administration \url{https://www.sba.gov/sites/default/files/advocacy/%bds_firm_size.xlsx
}} Though the data shows a lowering of exponent in lower-size part in contrast to higher-size part which could stand for a convexity of the distribution (Fig.~\ref{fig:FirmSizeDistrEvolv}), we could evaluate the tail exponent by linear regression, excluding both the lowest- and highest-size bins, because the highest bin has inappropriate upper limit for infinity. Figure~\ref{fig:TailExponentEvolv} shows clearly that the exponent decreases as time goes. 
Therefore, we can see that the tail exponent is evolving to lower, i.e. the distribution is flattening. We can give a reason in our approach as aforementioned: the variances of the age or the growth so increase that the variance of the size increases and the distribution of size flattens.

\section{Conclusion and discussion}

In this paper, we consider some special properties of distribution for the geometrically growing system (GSS) with pandemic, demographic and economic phenomena.

First, the distribution has a convexity in the lower-size part. It is not surprising, it represents only the modal (most probable) size, which is popular in almost distributions but absent in the power-law. In fact, the log-CS has additional concavity and singularity. In the above approximations the log-CS or log-normal, both or either, dominate over the power-law. This means that the demographic, pandemic and economic distributions can be explained by the GSS properly. In fact, difference between the log-CS and the log-normal is not great in most cases and not important.
What matters is that both of them represent the distribution of GSS and the convexity appears commonly in both them. However, profile of those distributions may be changed if the distribution of parameters is deviated from the normal. For the log-CS or log-normal, there does never appear a divergence problem which happens for the pure power law with a certain exponent.

If new items are born with low size continuously flourishing, this convexity will get fainter and the distribution seems to be closer to the power-law. 
However, once the number of items in the system is saturated, the number of low-size items decreases in evolution, unless they are isolated from the ensemble and never grow, and a kind of roll-over in lower-size part grows. 
Until early stage of such a period, the distribution of GGS should be represented only by the log-normal while long after the saturation of the number of items, the log-CS fitting can become possible.

Second, while the parameters such as the age and the growth diffuse, the tail exponent lowers and the distribution is flattening in the evolution of system, which is often called the spectral hardening. As aforementioned, the slope of distribution depends on the variances of the parameters fully or partly. 
In many systems such as Brownian motion, the variance of parameters are growing with time. The diversity in economical actions and variance in economical growth are accelerating with the time.
The second law of the thermodynamics dictates only that the matter should spread out by the diffusion. However, the matter is collecting and agglomerating over the universe. Though, in the physical view, it can be explained by the gravitation and so on, but, on the statistical way, the geometrically growing system can act as a reverse machine that converts the diffusion in parametric space to the concentration in the distribution of size.

The flattening distribution in turn implies the enlargement of the relative ratio in size between the highest- and lowest-size items within the system. 
This could explain the urbanization, monopolization and so on. The urbanization proceeded in ancient times such as in ancient Rome. The urbanization can occur not only by migration due to economic and political reasons, but also by the stochastic nature of the growth itself, e.g., by different birth (or death) rate or involving all the former factors. 
If we would apply this property to wealth distribution, which has geometrically growing trend and follows a power-law, 
we could expect aggravation of the ``rich-get-richer'' process and the monopolization in the 
economic regime by nature if money begets money. 

It is interesting that the concentration might be compatible with or even driven by the diffusion. The growth in GSS could give rise to the diffusion in parametric space, which in turn leads to the centralization in matter space.  
The ``rich-get-richer'' phenomenon does never imply that the rank in system should be fixed, that is, the richest or biggest one could keep their first rank naturally. The rank could be determined by the growth rate, the diversity of which can be changed with time. The richest job or biggest city have been alternating with era, as we have seen. 

We can find properties of the geometrically growing system in many other phenomena. We wish that our approach could contribute to analyze the problems.


\section*{Conflict of interest}
The author has no conflicts to disclose.

\section*{Data availability}
Data used in this paper are available at the website addresses indicated or by corresponding with the author.

\setcounter{equation}{0}
\setcounter{figure}{0}



\end{document}